\documentclass[useAMS,usenatbib]{mnras}
\usepackage[utf8]{inputenc}
\usepackage{natbib}
\usepackage{graphicx}
\usepackage{amsmath}
\usepackage{hyperref}
\usepackage{amssymb}

\newcommand{\eq}[1]{\begin{equation}  #1 \end{equation}}

\newcommand{\br}[1]{\left( #1 \right)}

\newcommand{\dd}{{\rm d}}

\def\apj{ApJ}
\def\apjl{ApJ}
\def\apjs{ApJS}

\def\mnras{MNRAS}
\def\nat{Nature}

\def\prl{Phys.~Rev.~L}
\def\pre{Phys.~Rev.~E}

\title[Iterative mean-field to spherical collapse]{Iterative mean-field approach to the spherical collapse of dark matter halos}

\author[Xun Shi]{Xun Shi$^{1}$\thanks{E-mail:
xun@ynu.edu.cn}\\
$^{1}$South-Western Institute for Astronomy Research (SWIFAR), Yunnan
University, 650500 Kunming, P. R. China\\
}

\begin{document}

\maketitle

\label{firstpage}
\begin{abstract}
Gravitational collapse of dark matter overdensities leads to the formation of dark matter halos which embed galaxies and galaxy clusters. An intriguing feature of dark matter halos is that their density profiles closely follow a universal form irrespective of the initial condition or the corresponding growth history. This represents a class of dynamical systems with emergent universalities. We propose an ``iterative mean-field approach'' to compute the solutions of the gravitational collapse dynamics. This approach iteratively searches for the evolution of the interaction field $\phi(t)$ -- in this case the enclosed mass profile $M(r,t)$ -- that is consistent with the dynamics, thus that $\phi(t)$ is the fix-point of the iterative mapping, $\mathcal{H}(\phi) = \phi$. The formalism replaces the N-body interactions with one-body interactions with the coarse-grained interaction field, and thus shares the spirit of the mean-field theory in statistical physics. This ``iterative mean-field approach'' combines the versatility of numerical simulations and the comprehensiveness of analytical solutions, and is particularly powerful in searching for and understanding intermediate asymptotic states in a wide range of dynamical systems where the solutions can not be obtained through the traditional self-similar analysis.

\end{abstract}


\section{Introduction} 
Large, gravitationally bound structures in the Universe, such as galaxies and clusters of galaxies, form from accumulations of matter collapsing under gravitational attraction. In the currently well-established Cold Dark Matter model, dark matter halos form during the gravitational collapse in the centers of the gravitational field of the primordial density fluctuations. They embed the galaxies and clusters of galaxies and play a governing role in their formation and evolution. 

One remarkable property of the dark matter halos is that halos with masses across several orders of magnitude all exhibit relative universal density profiles -- the so-called NFW profile \citep{nfw96,nfw97} which describes a Z-shaped transition from a logarithmic slope of -1 in the center and -3 in the outskirts. This universal profile was discovered in N-body numerical simulations, and is confirmed by later simulations and observations. Some evidence of deviation from the NFW profile emerged in recent simulations with unprecedented dynamic ranges, e.g. the results of \citet{wang20} preferred an Einasto profile featuring a smoother change of slopes, \citet{ishiyama10}, \citet{angulo17} and \citet{delos22} showed that when the free-streaming scale is taken into consideration (i.e. when dark matter is not taken to be perfectly cold), the density profiles of the first, earth-mass halos have -1.5 inner slopes. Nevertheless, NFW-like density slopes emerge as these halos evolve \citep{ishiyama14, angulo17, ogiya18, delos22} as a pure result of non-linear gravitational dynamics. This universal density profile can be regarded as a posterchild example of self-organized emergent universality in a complex system, where both the microscopic interaction and the initial condition are simple and fully understood. However, in this apparently clean example, the origin of this emergent feature remains intriguing despite the many efforts to understand it \citep[e.g.][]{ascasibar04, lu06, dalal10, hjorth10, ludlow13, pontzen13, williams22}. 

To understand the origin of the universality, one needs to examine the mapping from the initial condition: the density distribution of the initial overdensity, to the resulting dark matter halo density profile at a late time. This means the necessity of considering a large number of initial density distributions to sample the space of the initial conditions. Thus, it is desirable to have a fast computational method to map from an initial condition to the halo density profile at a late time. 

The spherical collapse model provides a simplified framework for examining such a mapping. Taking advantage of the approximate spherical symmetry of the gravitational collapse of high-density peaks, the spherical collapse model \citep{gunn72} considers the evolution of spherical shells of matter surrounding a spherical top-hat over-density embedded in a homogeneous expanding universe: the gravitational attraction of the initially over-dense region causes the shells to deviate from the Hubble flow and collapse onto it. The trajectory of a dark matter shell can be computed analytically when its enclosed mass stays constant. However, after the dark matter shells cross each other during gravitational collapse, the enclosed mass is no longer conserved, and thus, the evolution of many shells must be considered at the same time in the halo-forming region.

Traditionally, two approaches can solve this problem. One is the numerical simulation which evolves the dark matter shells by computing the instantaneous gravitational interaction among them. The other is the self-similar approach \citep{fillmore84, bert85, shi16b} which is an analytical approach that assumes self-similar symmetry, and thus applies only to a self-similarly expanding universe and to power-law initial over-density profiles. Both these approaches have their limitations. 

\begin{figure}
     \centering
         \includegraphics[width=.45\textwidth]{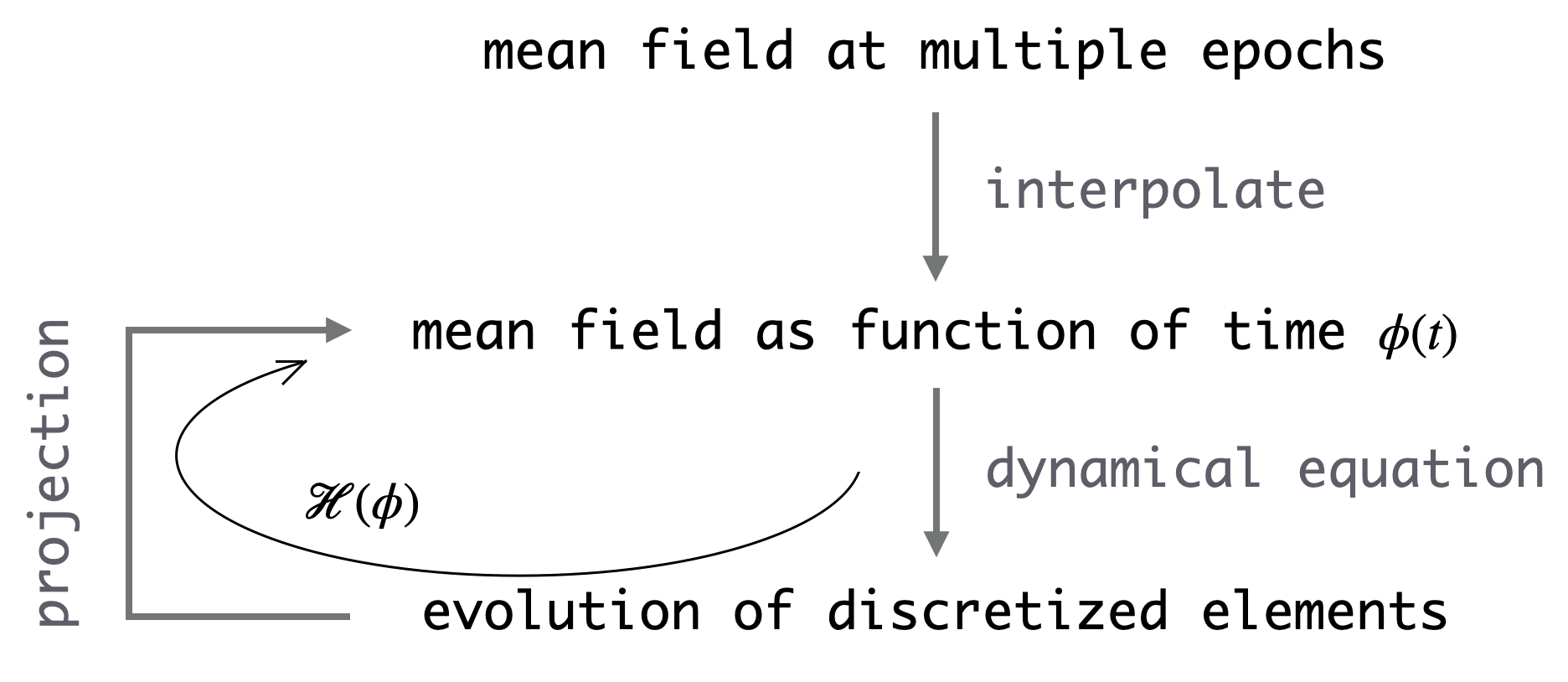} \\
       \caption{Summary of the iterative mean-field approach. The distribution of discretized elements lies in a state space with a dimension much higher than that of the mean field $\phi(t)$. The latter fully governs the evolution of the former under the mean-field approximation, and the former can be projected to update the latter, forming an iteration map $\mathcal{H}(\phi)$. The consistent solution $\phi(t)$ is the fixed point of the iteration map, i.e. $\mathcal{H}(\phi) = \phi$.}
     \label{fig:iteration}
\end{figure}

We notice that the desired quantity here -- the density profile of the dark matter halo, or equivalently the distribution of its gravitational potential, can be considered as a mean field of the dark matter shells. The dynamics of the dark matter shells are fully determined by the evolution of this mean field, i.e. this one-body interaction between the density profile and a dark matter shell can replace the N-body interaction among the dark matter shells to evolve the dynamics. On the other hand, the density profile at any time can be computed from the spatial distribution of the dark matter shells. This allows us to design a new iterative approach (Fig.\;\ref{fig:iteration}) -- the `iterative mean-field approach' to map the initial condition to the later time halo density profile that we are interested in.

\section{Spherical collapse dynamics}
In the standard $\Lambda$CDM cosmology, the evolution of a spherical overdense region is purely determined by its self-gravity.  When we divide this region into $N$ spherical shells, the trajectory of a dark matter shell $r_j(t)$ is governed by \citep[e.g.][]{mo_book}
\eq{
\frac{\dd^2 r_j}{\dd t^2} =  - \frac{GM(r_j,t)}{r_j^2} + \frac{\Lambda r_j}{3}  
\label{eq:d2rdt2}
}
for $j = 1,..., N$, where $\Lambda$ is the cosmological constant. 

The initial condition, if chosen at a sufficiently early time $t_i$ when gravitational collapse has not yet occurred, can be specified by the overdensity enclosed by the shell $\delta(r_j, t_i) \equiv \bar{\rho}(<r_j) / \bar{\rho}_{\rm m} - 1$, where $\bar{\rho}(<r_j)$ is the average density within the spherical shells, and $\bar{\rho}_{\rm m}(t)$ is the mean density of the universe. By definition, $M(r, t_i) = 4\pi \bar{\rho}_{\rm m} r_j(t_i)^3 (1 + \delta(r_j, t_i)) / 3$, and because $\delta(r_j, t_i)\ll 1$ at early time, the initial velocity of the dark matter shells follow the Hubble flow.  

As dictated by Eq.\;\ref{eq:d2rdt2}, a shell first expands with the universe at a rate only slightly smaller than the Hubble rate due to the small overdensity, but with time, the overdensity increases, and eventually the shell stops expending, turns around, and starts to collapse. 

For a typical overdense region whose mean density enclosed by a shell decreases monotonically with radius, a shell does not cross other shells until it starts to collapse. Thus, its enclosed mass keeps its initial value $M = M_i$ before the turn-around. Afterward, the mass enclosed by the shell is contributed by all shells currently at a radius smaller than the radius of this shell,

 \eq{
M(r_j, t) =  \sum_k m_k H\br{r_j(t) - r_k(t)}   \,,
\label{eq:summass} 
}
where $H$ is the Heaviside function.

In practice, we adopt the dimensionless form of the dynamical equations for simplicity (Appendix.\;\ref{app:dimensionless}).

\section{\label{sec:iter} The iterative mean-field approach}

The aim is to map the initial overdensity profile $\delta(r, t_i)$ to the late-time dark matter halo density profile, or equivalently, the enclosed mass profile $M(r, t)$ by solving the spherical collapse dynamical equation. Here, the enclosed mass profile is a representative of the mean field $\phi(t)$.

We design the iterative mean-field approach to this problem as follows.
First, we make an initial guess of the mass profile of the entire region at several epochs. Then, supposing that the mass profile evolves smoothly with time, we interpolate the guessed profiles in time to obtain the full function $M(r, t)$. 
Then, we iterate over the following two steps: 
\begin{enumerate}
     \item Compute the trajectories of all shells after their turnaround points using Eq.\ref{eq:d2rdt2} and the given mass profile $M(r, t)$. 
     \item From the shell trajectories, update $M(r, t)$ using Eq.\;\ref{eq:summass}.  
\end{enumerate}

The basic idea of the iterative mean-field approach is summarized in Fig.\;\ref{fig:iteration}. 
This approach allows us in step 1 to compute the evolution of each discretized element \textit{individually} using one numerical integration, i.e. it reduces the original N-body problem to a one-body problem. Compared to the basic form of an N-body simulation with $N^2$ interactions, this reduces the demand for numerical resolution in terms of the number of the discrete elements and the number of time steps, and thus improves computational speed. Also, this distinguishes the iterative approach from the basic form of numerical simulations where the interactions among the shells enter explicitly into the computation. The latter also means that the approach is free of the numerical instabilities that arise from the interactions of the discretized elements.

Note that, however, while the original form of an N-body simulation includes $N^2$ interactions among the particles, modern N-body simulations already use techniques e.g. the Particle-Mesh scheme to reduce the number of interactions and to boost the computational speed. These techniques also involve a smoothing of the force field like our iterative mean-field approach. In practice, the smoothing used in numerical simulations is typically only of a small degree. For example, in the case of Particle-Mesh, the resolution of the mesh controls the degree of the coarse-graining, and it is usually chosen so that the force field still represents the discrete N-body field. Nevertheless, in principle, the smoothing can be increased to further reduce discreteness effects. In this sense, the difference between modern numerical simulations and our iterative mean-field approach in computing an instantaneous force field is only quantitative, not qualitative. The qualitative difference lies in the fact that in our approach, the full history of the force field is also smoothed in the form of the evolution of the mean field. Therefore, not only individual forces at a given time can be computed independently of other discrete elements, but also the full trajectories of the discrete elements can be computed independently of each other at one iteration, i.e., our approach decomposes the dynamic problem into true one-body interactions. 

Another characteristic of the mean-field approach is that it is not sensitive to transient features. The dynamical evolution of the discretized elements usually involves the integration of the mean field over time, and the map from the state space of the discretized elements to the mean field is a projection from a high-dimensional space to a one-dimensional space that involves spatial averaging -- in other words, the two elementary steps in the iterative mean-field approach are both smoothing operations. This on one hand supports the convergence of the iteration, and on the other hand, makes the iterative mean-field approach suitable for understanding emergent universalities in dynamical systems.

\begin{figure}
     \centering
         \includegraphics[width=.38\textwidth]{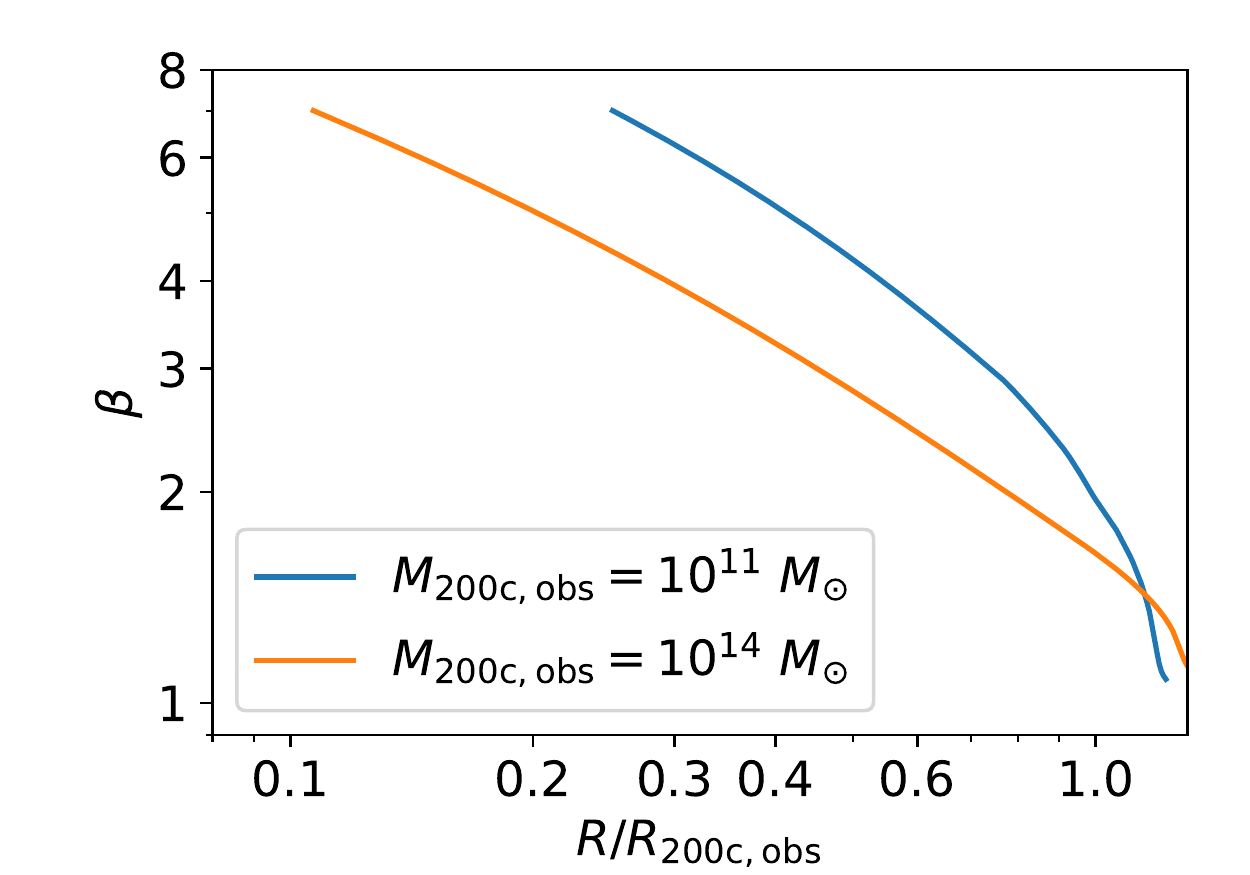} \\
         \includegraphics[width=.38\textwidth]{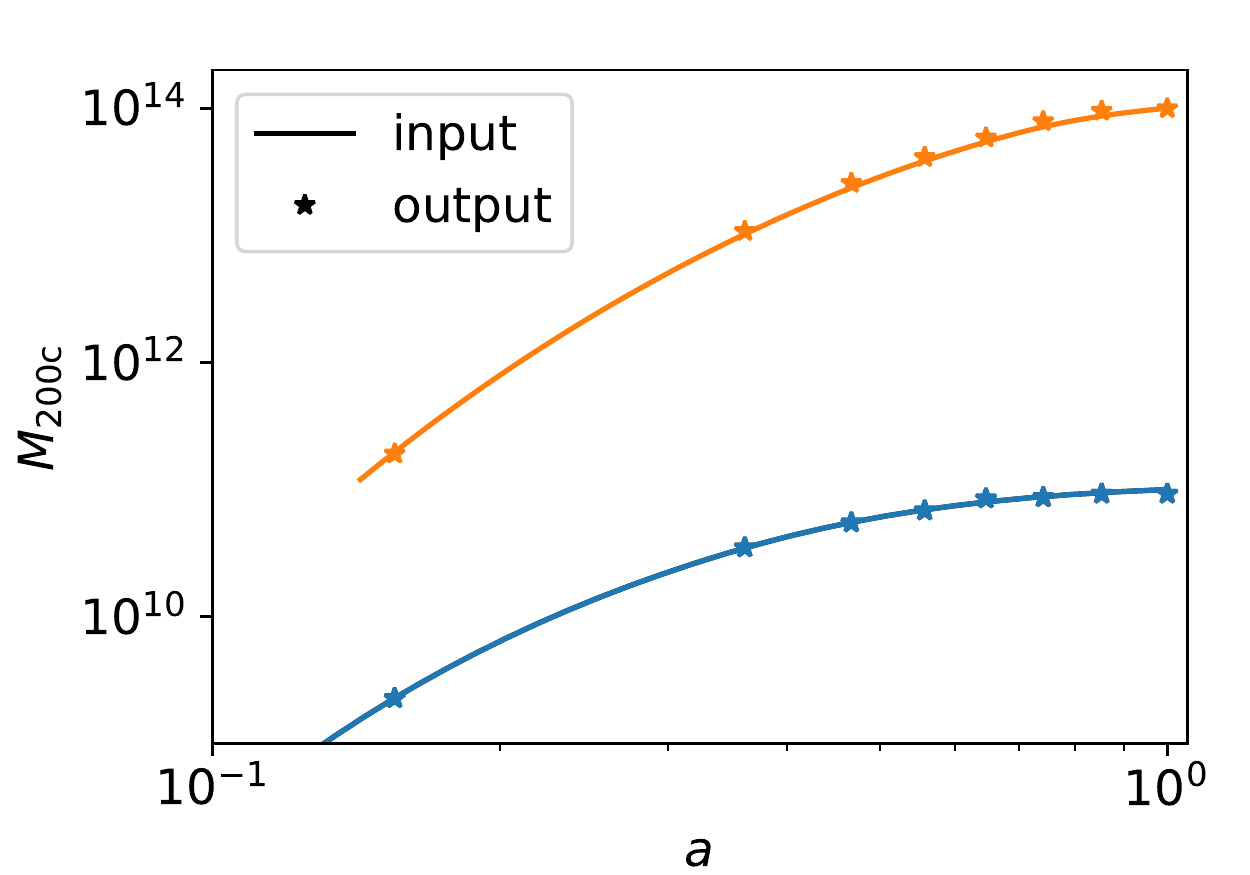} 
       \caption{Scaled initial overdensity profiles $\beta(r)$ (upper panel) derived from the
            average mass accretion histories described in terms of the mass within a radius enclosing an average density of 200 times the critical density of the universe $M_{\rm 200c}$ as a function of the scalee factor $a$ for $10^{11}$ and $10^{14}$ $M_{\odot}$ halos (lower panel). Spherical collapse results (stars in the upper panel) using the latter as an initial condition can reproduce the former (lines in the upper panel).} 
     \label{fig:MAH}
\end{figure}

\begin{figure}
     \centering
         \includegraphics[width=.4\textwidth]{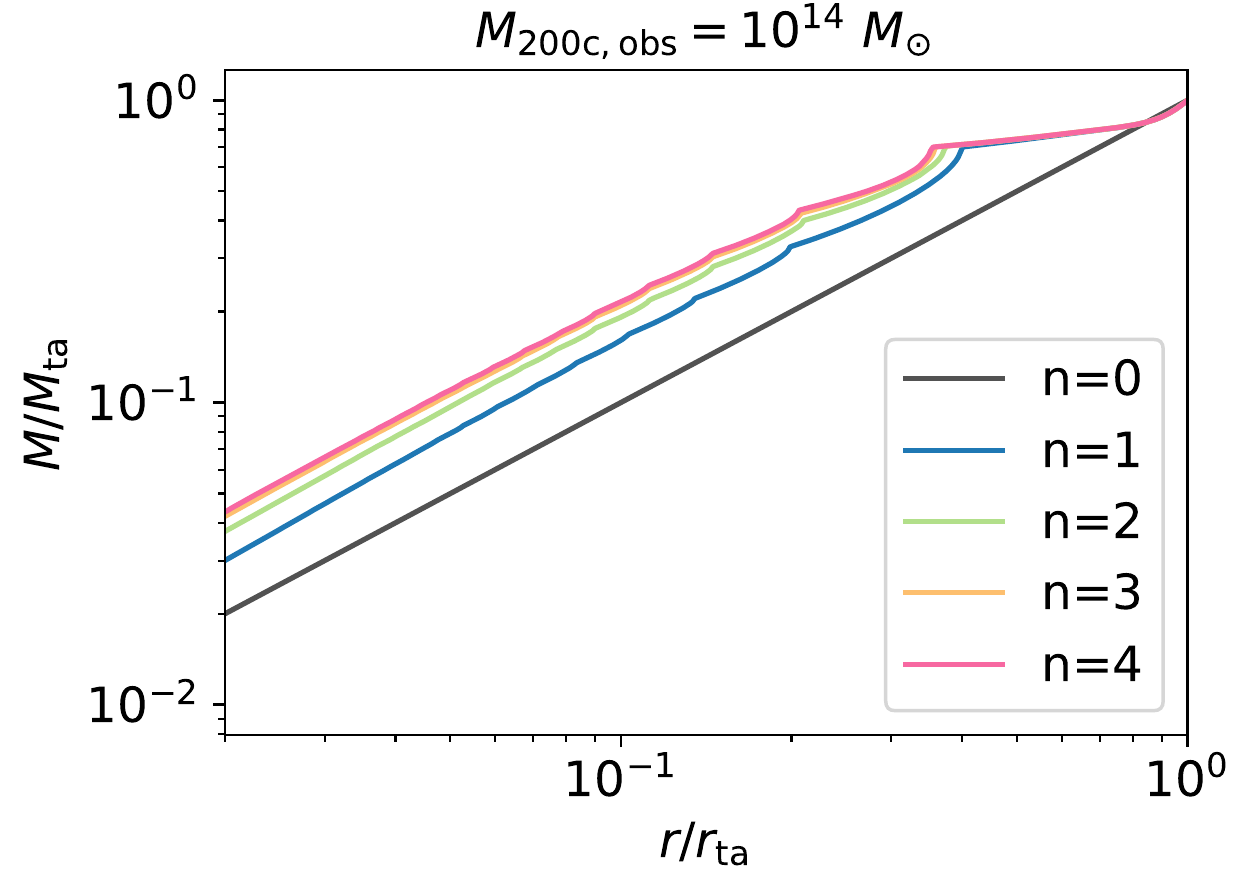} \\
       \caption{Quick convergence of the mass profile with the number of iterations $n$.}
     \label{fig:convergence}
\end{figure}

\begin{figure}
     \centering
         \includegraphics[width=.42\textwidth]{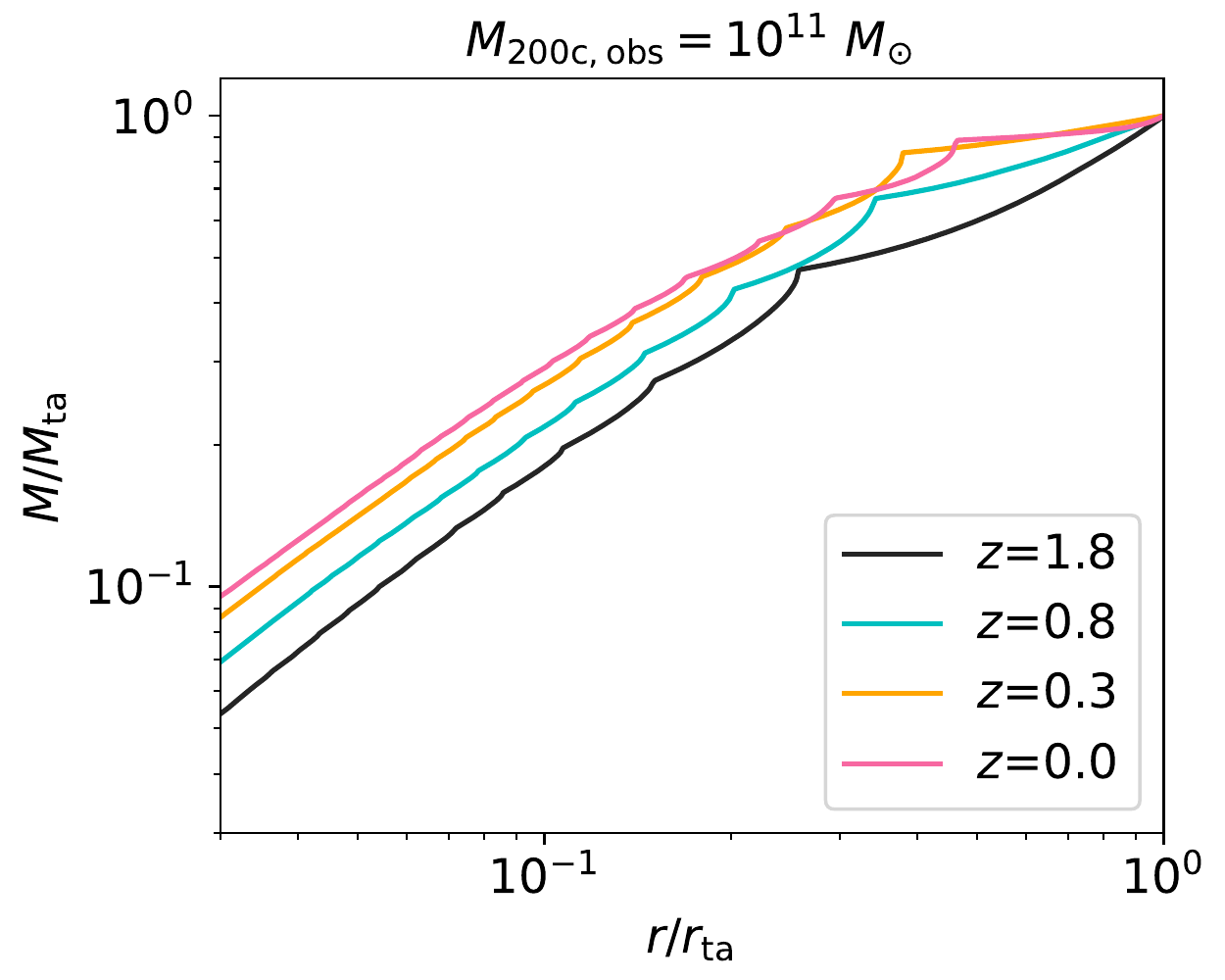} \\
         \includegraphics[width=.42\textwidth]{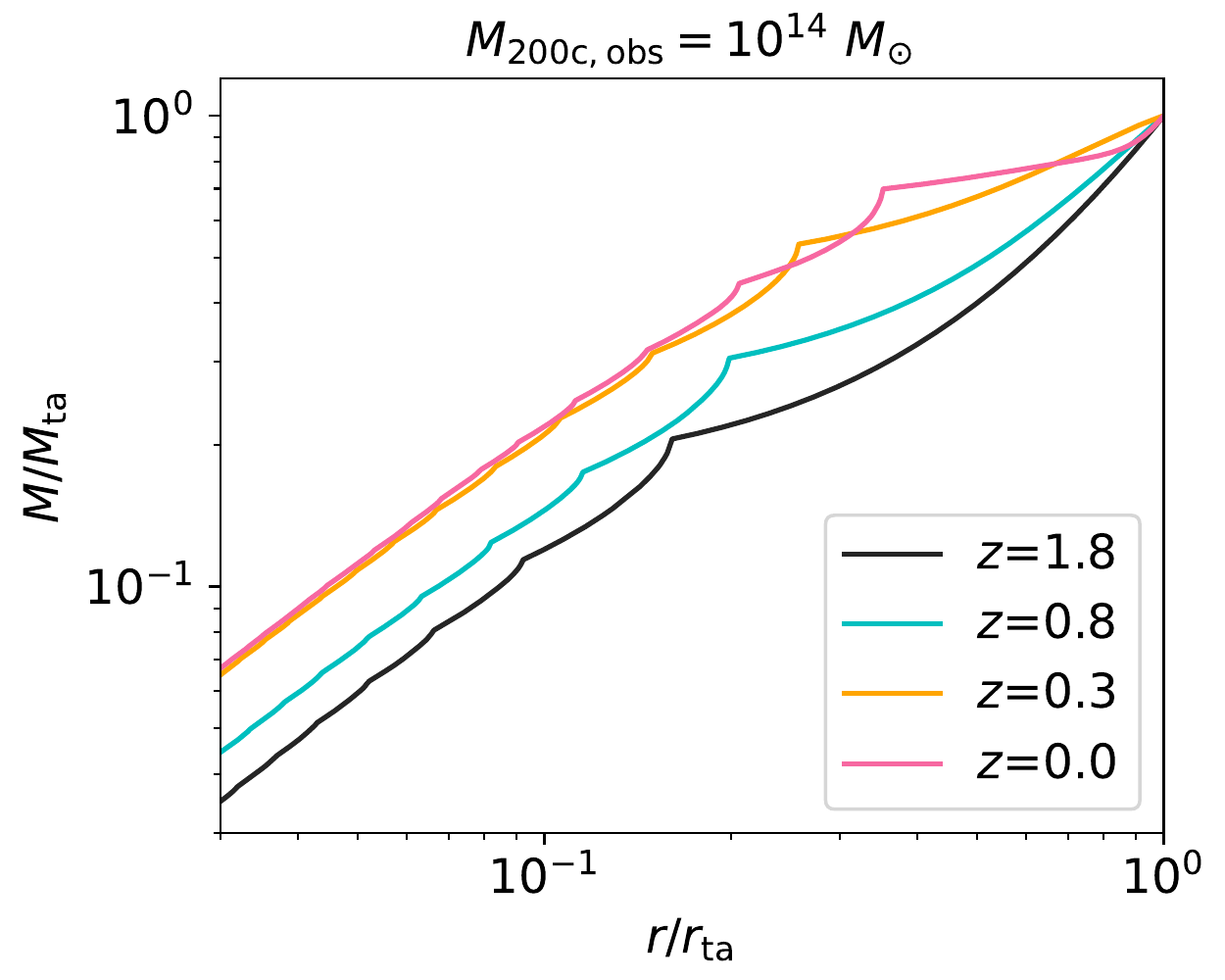}
       \caption{Scaled enclosed mass profiles at different redshifts for the $10^{11}$ $M_{\odot}$ (upper panel) and $10^{14}$ $M_{\odot}$ (lower panel) halos. Different mass accretion histories lead to their different mass profile evolutions.}
     \label{fig:massprof}
\end{figure}

\section{Results for halos with different mass accretion history}
As a demonstration, we consider two sets of initial conditions given by the initial overdensity profiles shown in the lower panel of Fig.\;\ref{fig:MAH}. The initial condition is described in terms of the scaled overdensity $\beta  = 5 \times 2^{2/3} \delta / (9 a) / (\Omega_{\rm m,0}^{-1} - 1)^{1/3}$ (see Appendix.\;\ref{app:dimensionless}) as a function of the radius $r$ scaled to $r_{\rm 200c, obs}$, a typical radius of the dark matter halo that encloses a region with mean overdensity equal to $200$ times the critical density of the universe at the time of observation today.

As the initial guess of the mass profile evolution, we simply use a linear relation $M / M_{\rm ta} = r / r_{\rm ta}$ at all times. We have scaled the quantities with those at the turn-around radius $r_{\rm ta}$ which separates the outside region where the dark matter shells still expand with the universe and the inside region where the dark matter shells start to fall toward the center of the overdense region. The turn-around radius at a certain time can be computed as the radius of the dark matter shell whose velocity is zero at that time.

For our demonstration here, we work in a flat $\Lambda$CDM universe composed of matter and dark energy, with Hubble constant $H_0= 70$ km/s/Mpc, and the current mean matter density $\Omega_{\rm m,0} = 0.27$, where subcript `0' denotes the present time where the cosmic scale factor $a=1$. We start our computation at an early redshift $z = 3.4$ where the universe is deep in the matter dominated regime with a mean matter density $\Omega_{\rm m} = 0.99$. We compute the trajectories of each discretized shell by integrating Eq.\;\ref{eq:d2rdt2} with the guessed mass profile evolution as an input. After completing the integrations at $z=0$, we reconstruct the mass profiles $M(r, t_k)$ for a radial range of $r < r_{\rm ta}$ at a few time steps $t_k$ by summing over the masses of the shells locating within radius $r$ at the time $t_k$ (see  \citealt{shi16a} for more details). Interpolating the reconstructed $M(r, t_k)$ over time, we obtain an updated mass profile evolution $M(r, t)$.

Then we iterate on $M(r,t)$ by repeating this process. Fig.\;\ref{fig:convergence} shows the mass profile at the time of observation at $z=0$ after each iteration. The iteration converges rapidly, and after four iterations the mass profile at the halo outskirts reaches percent level convergence. This fast convergence validates the use of the iterative approach. 
One condition that contributes to the fast convergence in the spherical collapse problem is the separation of time scales: the spherical collapse dynamical scale is much smaller than both the time scale on which the mass accretion rate varies and the time scale on which the background universe expansion rate varies. Short-time features on the mass accretion history or those on the universe expansion rate, if they exist, would correspond to transient features in the solution that would require more computational cost to capture with the iterative approach. 

Instead of solving the spherical collapse as an initial value problem given the initial overdensity profile (lower panel of Fig.\;\ref{fig:MAH}), it is sometimes desirable to compute mass profiles for halos with a specific mass accretion history (upper panel of Fig.\;\ref{fig:MAH}). Here, we actually derived the former from the latter using another iterative approach. We first use the top-hat spherical collapse trajectory obtained from Eq.\;\ref{eq:dydI2} to map the cosmic scale factor $a$ to a $\beta$ value that collapses at that time. 
Together with $M \propto r^3$, this gives an approximate initial overdensity profile. After obtaining the solutions using this as the initial condition, one can correct the initial overdensity profile by comparing the input and output mass accretion histories. Typically, by iterating once one can get the output mass accretion history consistent with the input one, as shown by the upper panel of Fig.\;\ref{fig:MAH}. 

The mass accretion histories we used correspond to the average mass accretion histories for $10^{11}$ and $10^{14}$ $M_{\odot}$ halos as given by \citet{zhao09}, see upper panel of Fig.\;\ref{fig:MAH}. The average mass accretion history for the more massive $10^{14}$ $M_{\odot}$ halos is steeper than that for the $10^{11}$ $M_{\odot}$ halos, especially at low redshifts where the cosmic scale factor approaches unity. This corresponds to a flatter, more extended overdensity profile of the initial density peak for the more massive halos (lower panel of Fig.\;\ref{fig:MAH}). This difference in the mass accretion history is imprinted in the resulting evolution of the enclosed mass profile (Fig.\;\ref{fig:massprof}) or the density profile which is the derivative of it. Even after scaling to the  turn-around values (thus subtracting the direct influence of the mass accretion history on the normalization of the mass profile), the mass profiles evolve differently in the two cases. 

Our results serve as a demonstration of the iterative mean-field approach. They cannot yet shed light on the universality of the NFW profile because of the neglection of angular momentum and the triaxiality of the dark matter shells. These, however, can be taken into account by generalizing the spherical collapse framework as \citet{lithwick11} has done to the self-similar spherical collapse. Applying the iterative mean-field approach to the generalized triaxial collapse framework is promising in helping solve the mystery.

The current spherical collapse results can already help understand the outskirts of the dark matter halos where radial motions dominate. Especially, the `splashback' feature \citep{adhikari14, diemer14} which corresponds to the outmost kink in the mass profiles (Fig.\;\ref{fig:massprof}), is being intensively studied as a promising physical boundary of dark matter halos \citep{more15, diemer20} which may have observation signals from gravitational lensing \citep{chang18_abb, contigiani19} and galaxy distributions in galaxy clusters (e.g. \cite{tomooka20, adhikari21_abb,shin21_abb} and references therein). Our formalism can establish a direct, quantitative link between halo mass accretion history and the location of the splashback feature. For example, as Fig.\;\ref{fig:massprof} shows, the splashback location is farther away from the turn-around location and evolves faster with time for the more massive halos. This is a result of the larger mass accretion rate of these halos at the current time, an effect that has been qualitatively demonstrated in our previous paper \citep{shi16a}.

\section{Discussion}
\subsection{Comparison to analytical approach and numerical simulation}
\begin{figure}
     \centering
         \includegraphics[width=.47\textwidth]{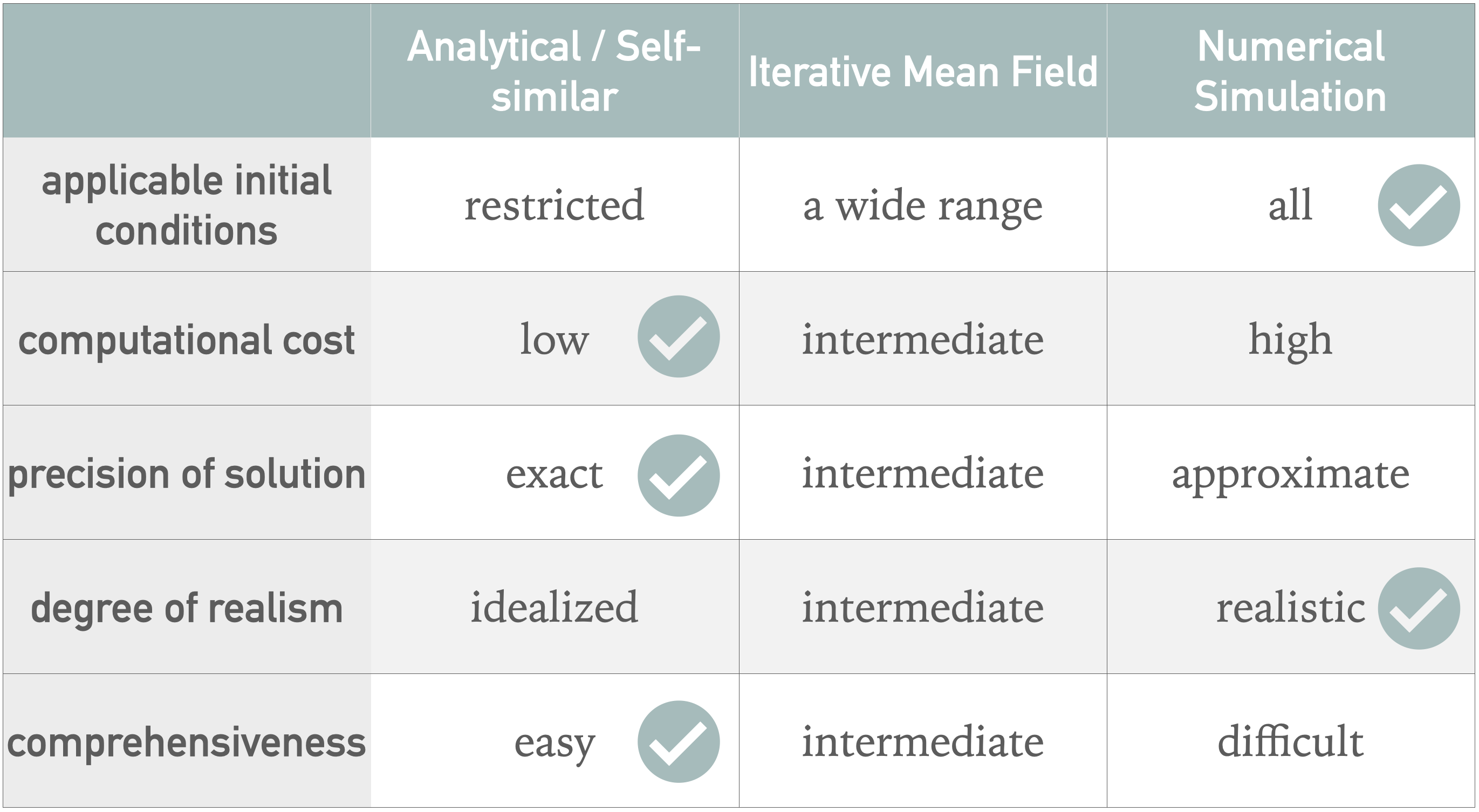} \\
       \caption{Comparison of approaches for solving spherical collapse dynamics. Both the analytical approach and the numerical simulation have advantages labeled by the tick icons. The iterative mean-field approach lies between these two approaches in terms of all properties listed here.}
     \label{fig:compare}
\end{figure}

The iterative mean-field approach has properties that lie between those of the analytical approach and the numerical simulations, as summarized by Fig.\;\ref{fig:compare}.  For a non-linear dynamical problem such as spherical collapse, the canonical approach is numerical simulations which can cope with all initial conditions and give realistic results with a lot of details. However, their results are approximate in the sense that they depend on the numerical resolution, and that they may suffer from numerical instabilities arising from the interactions among the discretized elements. Their results are too detailed to be easily comprehensible, and their computational cost is high. 

The analytical approach is highly complementary to numerical simulations. For spherical collapse, the analytical approach that can give the dark matter halo density profiles is the self-similar approach. It applies only to very restricted initial conditions which also means that its results are highly idealized, but it is easy to compute and yields exact solutions which are easy to interpret. 

The iterative mean-field approach, like numerical simulations, works on discretized elements but features a reduction of the original N-body interactions among the discretized elements to one-body interactions of each discretized element with an evolving mean field. Thus, it is numerically very different from, and complementary to the numerical simulations. An instantaneous mean field can be regarded as the result of an extreme form of coarse-graining of the force field that is possible yet usually not applied in numerical simulations. Guessing the time evolution of the mean field and using iterations to let it converge, on the other hand, is unique to the iterative mean-field approach and is what offers it a unique perspective on the dynamical behavior of the system. Compared to the self-similar approach, the iterative mean-field approach is in principle applicable to all initial conditions but is most suitable for initial conditions that assure scale separation i.e. that are sufficiently smooth. For this wide range of smooth initial conditions, convergence can be achieved with a relatively small numerical resolution in space and time.

\subsection{As an extension of the self-similar approach}
Self-similar solutions often represent the intermediate asymptotics of the dynamical behavior \citep{zeldovich67, barenblatt96}. This means that these solutions are not simply particular cases, but rather, they describe the asymptotic behavior of a wide class of situations. In the sequence of studies starting with \citet{goldenfeld89, goldenfeld90, goldenfeld91, chen94, chen95}, a deep connection between the self-similar solutions and the renormalization group is uncovered: the self-similar solutions are found to be fixed points of the renormalization group transformation, and that an iterative use of the latter is an efficient way of finding the former.  
The intermediate asymptotic property and the connection to the renormalization group make the self-similar approach attractive in helping understand emergent universal behaviors of dynamical systems.    

However, in some cases like that of the spherical collapse, self-similarity is too restrictive, and emergent universality still occurs in non-self-similar conditions. In such cases, the iterative mean-field approach may work as an extension of the self-similar approach. As is shown with the spherical collapse example, it combines well with scaling analysis and welcomes dimensionless forms of dynamical equations that one can typically obtain using Buckingham's $\Pi$-theorem \citep{buckingham14}. 

For the spherical collapse problem, the iterative mean-field approach is practically developed as an extension of the self-similar approach. In the self-similar spherical collapse model \citep{fillmore84,bert85}, an iteration between the mass profile and the dark matter shell trajectory has been used to obtain the mass profile $M(r / r_{\rm ta})$ for self-similar initial conditions. We noticed that this iterative approach is not limited to self-similar solutions, that when we allow a time dependence of the mass profile $M(r / r_{\rm ta}, t)$, the generalized iterative approach can solve the spherical collapse dynamics for non-self-similar initial conditions, and hence the development of the iterative mean-field approach. The same may apply to other systems where the self-similar solutions are highly appreciated but one is also interested in behaviors in not-so-self-similar conditions.

\section{Conclusion}
\label{sec:conclusion}
The non-linear formation of dark matter halos via gravitational collapse can be considered as a posterchild example of complex non-linear dynamical systems with emergent universal features. To understand the emergent feature here -- the universal density profile, one needs to examine the mapping from the initial condition to the late-time dark matter halo density profile for a wide range of initial conditions.   

We presented a new computational method -- the iterative mean-field approach for this purpose. 
This method singles out the `mean field' -- the distribution of the gravitational field as a key quantity in the dynamics, makes an initial guess on its evolution, and iterates upon it using the distribution of the dark matter elements evolved by the dynamics.    
The iteration converges rapidly with the number of iterations, yielding the desired late-time dark matter halo density profile corresponding to a certain initial condition.  

The iterative mean-field approach is highly complementary to both analytical methods and numerical simulations. Combining the mean-field approximation with an iterative scheme, it decomposes the original N-body problem into one-body problems by letting the discretized elements evolve independently of each other. Thus, it is numerically very different from the numerical simulations and offers a new perspective on the dynamical system. The new method can apply to a wide range of initial conditions and is thus more flexible than analytical methods.  

As a demonstration, we applied the new approach to a simplified framework of dark matter gravitational collapse -- the spherical collapse framework, and computed the evolution of density profiles for two sets of initial conditions representing dark matter halos with different mass accretion histories. The new approach establishes a direct, quantitative link between halo mass accretion history and the corresponding density profile evolution. Within the current spherical collapse framework, the new approach can enable a detailed study of the dependence of the splashback location on the halo mass accretion history. If generalized to triaxial collapse with angular momentum, it is promising to shed light on the physical origin of the universal NFW profile.

It is possible that the iterative mean-field approach can apply to a wide range of dynamical systems, and even act as a generalized method for solving ODEs and PDEs to look for low-dimensional universalities. The easy combination of the iterative mean-field approach with scaling analysis also makes it an interesting possibility of extending the idea of self-similar analysis to not quite self-similar conditions. We expect it to contribute to the endeavor of finding and understanding emergent features in complex dynamical systems.\\

\noindent\textbf{Data Availability}
No datasets were generated or analysed during the current study.\\

\noindent\textbf{Acknowledgments}
\noindent XS thanks GuangXing Li and the participants of the splashback workshop, especially Neal Dalal, for helpful discussions, and the referee for the helpful report. We also acknowledge Eiichiro Komatsu for the encouragement to give a lecture on spherical collapse in the MPA staff/postdoc cosmology lecture series which stimulated this work. This work is supported by NSFC grant No. 11973036.

\bibliographystyle{mnras}
\providecommand{\noopsort}[1]{}\providecommand{\singleletter}[1]{#1}%

\appendix

\section{Dimensionlesss equations and initial conditions of spherical collapse}
\label{app:dimensionless}
In a flat $\Lambda$CDM cosmology, the dynamical equation Eq.\;\ref{eq:d2rdt2} can be casted into a simple dimensionless form  \citep{shi16a}, 
\eq{
\label{eq:dyn}
\frac{\dd^2 y}{\dd I^2} =
-\frac{M}{2M_i} \frac{1}{y^2} + y \,,
}
with scaled radius $y(t) = w^{\frac{1}{3}} r(t) / r(t_i)$, and scaled time $I = \br{1-\Omega_{\rm m0}}^{\frac{1}{2}} H_0 t$, where the scaled cosmological constant $w = \Lambda / 8\pi G \bar{\rho}_{\rm m,0} = \Omega_{\rm m,0}^{-1} - 1$.

The initial velocities, as well as the velocities before shell crossing, can be directly obtained by integrating Eq.\;\ref{eq:dyn} replacing $M / M_i = 1$, which yields
\eq{
\label{eq:dydI2}
\br{\frac{\dd y}{\dd I}}^2 = 
y^{-1} + y^2 - 3\beta/2^{\frac{2}{3}} \,.
}
The integration constant $\beta$ is related to the curvature $K$ of the enclosed region (integration constant of Eq.\;\ref{eq:d2rdt2}) as (see \cite{shi16a} for details)
\eq{
     \beta = \frac{2^{2/3}(1 + w)K}{3w^{1/3} H_0^2 R^2} , 
}
where $R = r/a$ is the comoving radius of the region. We shall use this dimensionless $\beta$ parameter to indicate the initial overdensity $\delta$ since the two quantities are linearly related to each other, with a positive $\beta$ related to positive overdensities. For linear perturbations in the matter-dominated regime, one can derive an explicit linear relation between $K$ and $\delta$ \citep[e.g.][]{wagner15}, which gives $3 w^{1/3}\beta / 2^{2/3} = 5 \delta / (3 a)$. It is advantageous to use $\beta$ because it does not rely on the specific choice of the initial time. Moreover, as one can infer from the properties of Eq.\;\ref{eq:dydI2}, the $\beta$ value alone determines the trajectory of the shell expressed in the scaled quantities. The larger the $\beta$ value, the earlier the turnaround and collapse will occur, and only regions with $\beta > 1$ will ever collapse.

For an initial overdensity profile expressed in terms of $\beta(R)$ (lower panel of Fig.\;\ref{fig:MAH}), we use Eq.\;\ref{eq:dydI2} to evolve 
each shell directly to its turnaround point, and then switch to Eq.\;\ref{eq:dyn} and start the iteration. 

\end{document}